\documentclass[aps,amsfonts,amnssymb,amsmath,notitlepage,twocolumn,reprint,showpacs]{revtex4-1}

\usepackage{graphicx}

\newcommand*\mylabel[1]{\label{#1}}
\newcommand*\myref[1]{(\ref{#1})}

\begin{document}
\title{Transitions in the Stock Markets of the US, UK, and Germany}
\title{Transitions in the Stock Markets of the US, UK, and Germany}
\author{Matthias Raddant}
\affiliation{Institute for the World Economy \\
Kiellinie 66, Kiel 24105, Germany,\\ and\\
Department of Economics, Kiel University\\
Olshausenstr. 40, 24118 Kiel, Germany,\\
matthias.raddant@ifw-kiel.de, } 
\author{Friedrich Wagner}
\affiliation{Institute of Theoretical Physics, Kiel University\\
Leibnizstr. 19, Kiel 24098, Germany, 
wagner@theo-physik.uni-kiel.de}

\begin{abstract}
In an analysis of the US, the UK, and the German stock market we find a change in the behavior based on the stock's beta values.
Before 2006 risky trades were concentrated on stocks in the IT and technology sector. Afterwards risky trading takes place for stocks from the financial sector. We show that an agent-based model can reproduce these changes. We further show that the initial impulse for the transition might stem from the increase of high frequency trading at that time.\\ \\
\noindent
An earlier version of this paper is available as Kiel Working Paper No. 1979.  

\end{abstract}


\maketitle

\section{Introduction}
In this paper we analyze transitions in the structure of the US, the UK, and the German
stock market. In particular we observe a phase of dominance of IT oriented stocks followed by a transition 
period that leads to a dominance of the financial sector.

The analysis of the differences in the returns of stocks have long been dominated by the discussions around different versions of 
a CAPM model \cite{lintner,sharpe}.
The original version of the CAPM is in fact a one factor model, which postulates that the
returns $r_i$ of the stocks should be governed by the market return $r_M$ and only differ by
 an idiosyncratic component $\beta_i$ for each stock $i$, such that
\begin{eqnarray} \mylabel{capm}
r_i(t) =  \beta_i r_{M}(t) + \epsilon_i(t) .
\end{eqnarray}
Hence, stocks differ by the amount of volatility with respect to the market, and economic
rational necessitates that higher stock volatility is compensated by higher absolute returns (additionally eq. \mylabel{capm} may incorporate the risk free interest rate).
Empirical tests of this model had rather mixed results and have let to two conclusions: 
More factors are needed to explain the variation of stock returns. The widely used Fama and 
French \cite{famafrench} model for example is a three-factor model that incorporates firm size and book-to-market ratio. 
The second conclusion was that beta values are not constant but time-varying, see \cite{tvcapm}. 
The reasons for the variability of the betas are manifold. They could change due to microeconomic
factors, the business environment, macroeconomic factors, or due to changes of 
expectations, see, e.g., \cite{bos,adcock,harvey}. Also models that assume a first-order auto-regressive process have been suggested, see \cite{bodurtha}.

Our approach to identify different states of a stock market consists in an analysis 
of a covariance matrix, similar to \cite{munnix}, and of the transaction volumes, like in \cite{preis}.
The properties of the covariance matrix of asset returns depend on the time horizon $T$ in
which they are determined. For short $T$ in the order of months they are rather
volatile, and partly mirror economic and political changes \cite{raddant,kenn}. 
 \cite{beile} for 
example argue that correlations increase in times of crisis, which has profound implication for
portfolio choice and hedging of risks. 

For large $T$ in the order of several years, a principal
component analysis \cite{lalo,plerou,alfa} of the correlation matrix is possible. The 
component with the largest eigenvalue can be interpreted as the market. The $\beta$ coefficients
are proportional to the corresponding eigenvector. 

In order to detect changes in stock betas we use time windows of less than 4 years
and a rather large numbers $N$ of stocks for different markets. In this case the principal component is well
separated from the rest. Within the assumption of market dominance motivated by implementing
eq. \myref{capm} with a stochastic volatility model (SVM), one can determine the  $\beta$ 
coefficients. A problem may be the statistical accuracy, which could be of order $\sqrt{N/T}$
as suggested by random matrix theory \cite{marc}. However, a Monte Carlo simulation shows that the errors for the
$\beta$ are in the order of $\sqrt{2/T}$.

This paper consists in a substantial extension of the research presented in \cite{wagner1},
with respect both to methods and data. 
The paper is organized as follows: In section \ref{sec:methods} we briefly describe the data sets 
before we describe the methodology to analyze the covariance matrix and the distribution of the stock returns.
In section \ref{monte} we present a Monte Carlo study for the error estimate on $\beta$. 
After this we show the transitions in the markets and introduce a sector specific risk measure. In section \ref{sec:explain}
we present a model that can replicate the transitions and we discuss whether the cause of the changes is an internal or external one. Section \ref{sec:conc} concludes.

\section{Materials and methods}\label{sec:methods}
\subsection{The data sets \label{data}}
\begin{table}
\begin{tabular}{r |c c c}
  &  US   & UK   &  Germany \\
	&  S\&P500   & FTSE350 & CDAX \\
	\hline
period & 1995--2013 & 1997--2013 & 1999--2013 \\
T     & 4782   & 4294   & 3691 \\
N     & 356    & 132    & 78   \\
 \hline \hline
\textit{sector }\\
Energy & 32  & 5  & 0 \\
Materials & 23  & 7  & 7 \\
Industrials & 51 & 30 & 25 \\
Cons. Discr.& 56 & 26  & 12 \\
Cons. Staples & 35 & 10 & 5 \\
Health & 32 & 5 & 10 \\
Financial & 60 & 37 & 7 \\
Technology/IT & 37 & 4 & 7 \\
Telecom. & 3 & 4 & 3 \\
Utilities & 27 & 4 & 2 \\
\end{tabular}
\caption{Summary statistics of the data sets}\label{tab:data}
\end{table}

For our analysis we use data from Thompson Reuters on the closing price of stocks which were continuously traded with sufficient volume
throughout the sample period and had a meaningful market capitalization \footnote{We excluded stocks which price behavior or market capitalization 
showed similarities to penny stocks, or which were exempt from trading or traded with negligible amounts for more than 10 days, or for which the trading 
volume was negligible for more than 8\% of the total trading days.}.
For the US we choose stocks which are part of the S\&P500 stock index. For the UK the stocks in our
sample are listed in the FTSE350, the German stocks are all part of the CDAX (and are with very few exemptions also listed in the MDAX, SDAX or DAX30).
The size of the US market allows us to collect a time series corresponding to 20 years of data. For the European markets it is
not possible to analyze a quite as long time horizon, since not enough stocks have been traded for such a time span. We have
collected the sector classification of the firms, using the GICS classification for the US market and the (for our purposes
practically identical) TRBC classification from Thompson Reuters for the European markets. Table \ref{tab:data} summarizes the data sets and the sector information.

\subsection{Analysis Method for Correlations \label{method} }

Stock markets can be analyzed by the study of the correlation between the returns
of the participating firms. The $N$ firms are indexed by $i=1,\cdots,N$. A
 return $r_i$ is given by the log of the price ratio between consecutive days.
The returns are normalized by
$\sum_{\tau=1,T_0} \sum_{i=1,N} r_i^2(\tau)=NT_0$. $\tau$ denotes the days in one time window.
 For the covariance matrix $C$ we
consider time windows  of size $T$ centered at time $t$. $C$ is given by
\begin{equation} \mylabel{aa1}
C_{ij}(t)=\left <r_ir_j\right >_{T,t}
\end{equation}
 with the abbreviation for the time average
\begin{equation} \mylabel{aa2}
\left <A \right >_{T,t}=\frac{1}{T} \sum_{\tau=t-T/2 }^{\tau<t+T/2 }A(\tau)
\end{equation}
for any observable $A$. In eq. \myref{aa1} the small time averages $<r_i>$ are neglected.
When $C$ is derived from the returns of many stocks in a long time window $T\propto T_0$, one usually observes that the matrix $C$ 
has one large eigenvalue $\lambda_0$ in the order of $N$ with a corresponding eigenvector that we denote $\beta_i$.
All $\beta_i$ have the same sign and can be chosen positive. We normalize
by $\sum_i \beta_i^2=N$. The remaining 
eigenvalues are of order of 1. The first eigenvector can, for example within the framework of a principal component analysis, be interpreted 
 as the market. This means that this eigenvector can be interpreted as the weights of the single stocks within the market factor.
Hence, a market return $r_M$ can be defined by the 
the projection of $r$ on $\beta$
\begin{equation} \mylabel{a3}
r_M(\tau)=\frac{1}{N}\sum_i\beta_ir_i(\tau)
\end{equation}
Due to the relation
\begin{equation} \mylabel{a3a}
\beta_i=\frac{\displaystyle \left < r_i r_M \right >_{T_0,t}}{\displaystyle \left <r_M^2 \right >_{T_0,t}}
\end{equation}
the components of the leading eigenvector are $\beta$-coefficients in a 
CAPM approach (leaving out the risk-free interest rate). With $T=T_0$ we would have only one vector $\beta_i$ centered at time $(T_0/2)$.
A time dependence of $\beta$ can be achieved by using a
moderate time window $T$ (in the order of years). 

To derive meaningful $\beta$s we assume
that the return follows a stochastic volatility model (see, e.g., \citep{svm,svm2}): The
returns are the product of a noise factor and a slowly varying stochastic volatility
factor. The latter should be considered as constant over the window size $T$.
Then eq. \myref{aa1} corresponds to an average over the noise with a statistical
error depending on the properties of $r_i$. 

As a first example we consider
$r_i(\tau)=\gamma \eta_{i\tau}$ with an i.i.d. Gaussian noise $\eta$. For a finite
$T$ we obtain a Marcenko-Pastur spectrum \cite{marc} spread over an interval $\gamma^2(1\pm 2\sqrt{N/T})$
(instead of the degenerate eigenvalue $\gamma^2$ ).
For $N\sim 400$ a time window of only a few years would lead to  prohibitive large
uncertainty. However, this model cannot account for the occurrence of one large 
eigenvalue.

This can be reproduced by the second example with 
$r_i(\tau)=\gamma_i \eta_{\tau}$. In this model all stocks follow the market described
by  Gaussian noise. For $T \to \infty$ the covariance matrix $C$ has one eigenvalue
$\lambda_0=\sum_i\gamma_i^2$ with eigenvector $\beta_i \propto \gamma_i$ and
$N-1$ zero eigenvalues. At finite $T$ the eigenvectors and the zero eigenvalues are unchanged. 
$\lambda_0$ is multiplied with a $\chi^2$ distributed
number with mean $1$ and variance $2/T$. To describe the observed spectrum of small
eigenvalues we consider a second process that leads to an additional 
additive component $C_{1ij}$ in $C$. 

We assume market dominance in the sense that $\gamma^2$ is of order $N$ and
$(\gamma,C_1^k\gamma)$=$A_k\gamma^2$ with constants $A_k$ is of order 1.
Perturbation theory for large $N$, see the appendix, shows that $C_1$ does 
not change $\lambda_0$ and $\beta_i$ up to $1/N$ contributions. The remaining 
eigenvalues are strongly dependent on the noise. Only their sum is given
by trace(C)-$\lambda_0$. Neglecting
very small quantities, $<r_i>_{T,t}$  is a measure of the volatility $v^2$ in the window.
\begin{equation} \mylabel{a3b}
v^2(t)=\frac{1}{N}\left(\mbox{tr}(C)  -\sum_i \left<r_i\right>_{T,t}^2\right)
\end{equation}
$\lambda_0$ determines the size of the market return $<r_M^2>$ via
\begin{equation} \mylabel{a3c}
\left <r_M^2\right >_{T,t}=\frac{\lambda_0(t)}{N}
\end{equation}


\subsection{The shape parameter of the returns distribution}

In order to analyze changes in the distribution of the stock returns we estimate the tail 
parameter of its pdf $f(r)$. We characterize $f$
by a Pareto-Feller distribution \cite{pfeller}, where $f$ depends only on
$r^2$ and a finite $f(0)$. The two parameters are a scale parameter $r_0$ and a 
tail index $\alpha$. It is given by

\begin{equation} \mylabel{a3d}
f(r)\propto \left(1+\frac{r^2}{(\alpha -2)r_0^2}\right)^{-(\alpha+1)/2}
\end{equation}

Performing fits with limited statistics $\alpha$ and $r_0$ are strongly correlated.
Therefore we fix $r_0$ by the condition $r_0^2=E[r^2]$.

\section{Monte Carlo Simulations \label{monte} }

\begin{figure}
 \includegraphics[width=0.9\linewidth, bb = 105 265 490 575, clip=true]{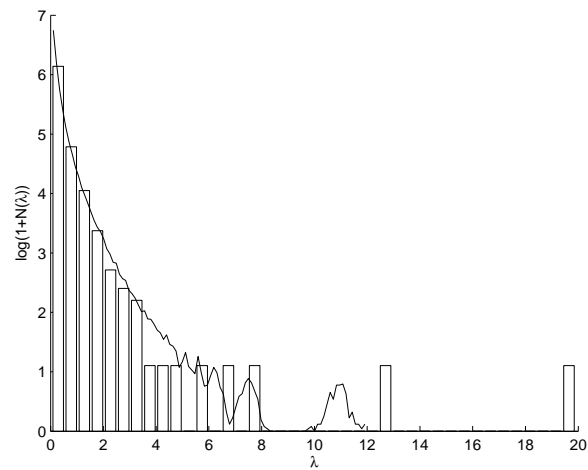}
   \caption{\label{figmc1} $\log(1+N(\lambda))$ for the empirical spectrum
            (histogram) $N(\lambda)$ from S\&P in 2004
            and simulated spectrum (line). For the simulation we used  
            log-normal distributed $\beta_i$ with mean 0.93, $\theta=0.26$
            and log normal distributed $\gamma_{1i}$ with $\gamma_0=0.865$.} 
\end{figure}

From the discussion in section \ref{method} we expect an error on the leading
eigenvalue $\lambda_0$ in the order of $\sqrt{2/T}$. This does not imply the same accuracy
for the eigenvector $\beta$. To estimate the size of errors in windows of several years, we
perform a Monte Carlo study based on a fairly general SVM. In a single window
we assume the following returns
\begin{equation} \mylabel{mc1}
r_i(t) = \sqrt{\theta}\; \gamma_{0i}\; \eta_t + \sqrt{1-\theta}\; \gamma_{1i}\; \eta_{it}
\end{equation}
Comparing equation \myref{mc1} with the CAPM definition \myref{capm} we see that
the first term corresponds to the market component with strength $\theta$
and the second term can be
interpreted as the idiosyncratic component due to trading activity for specific stocks.
For the i.i.d. noise factors we use $\eta_t ~ N(0,1)$ and
$\eta_{it} ~ N(0,1)$. We checked that a Laplacian noise 
for $\eta_{it}$ as suggested by
\cite{svm} does not change the result. The parameters $\gamma_{0i}$ and
$\gamma_{1i}$ are independent of time and normalized to $\sum_i\;\gamma_{ki}^2=N$.
For a window size $T \to \infty$ we get for $C$
\begin{equation} \mylabel{mc2}
C_{ij}=\theta \gamma_{0i}\gamma_{0j} +(1-\theta)\gamma_{1i}^2\; \delta_{ij}
\end{equation}
Using perturbation theory (see the appendix) the leading eigenvalue and its
eigenvector $\beta_i=\sqrt{N}f_i^0$ are given up to terms of order of $1/N$ by
\begin{equation} \mylabel{mc3}
\lambda_0=\theta (N-1)+1 \quad \mbox{and} \quad \beta_i=\gamma_{0i}
\end{equation}

\begin{table}
\begin{center}
\begin{tabular}{c |c c c c || c | c}
$N$   & log-norm. & Laplace   & normal & $\gamma_{1i}=1$ & T & log-norm. \\
\hline \hline
  100 &  0.109   & 0.109   & 0.108 & 0.125 &  500 & 0.131 \\
  200 &  0.107   & 0.109   & 0.110 & 0.124 & 1000 & 0.094 \\
  400 &  0.106   & 0.108   & 0.110 & 0.123 & 2000 & 0.066 \\
\end{tabular}
\end{center}
\caption{\label{tabmc} The left part of the table shows the average error (5\% confidence level)
         of $\beta$ at $T=750$ for various $N$ and models for $\gamma_1$.
         The right shows the error for $N=356$ and log-normally distributed  $\gamma_1$ for different $T$.}
\end{table}

We simulate the returns from eq. \myref{mc1} with given values for $\theta$, 
$\beta_i$ and $\gamma_{1i}$ in a finite window. From the eigenvectors of the
simulated covariance matrix we can estimate the statistical error on $\beta$
due to the finite $T$. For this we need reliable values of the input parameters.
The market strength $\theta$ follows from the well measured empirical $\lambda_0$.
For the input $\beta_i$ we use a log-normal distribution, which represents the 
observed spectrum very well. For $\gamma_{1i}$ we choose a normal, Laplace and 
log-normal distribution, which depend on two parameters. Since the mean of
$\gamma_{1}^2$ must be 1 only the mean $\gamma_{0}$ is a free 
parameter. 

\begin{figure}
  \includegraphics[width=0.9\linewidth, bb= 101 262 488 568, clip=true]{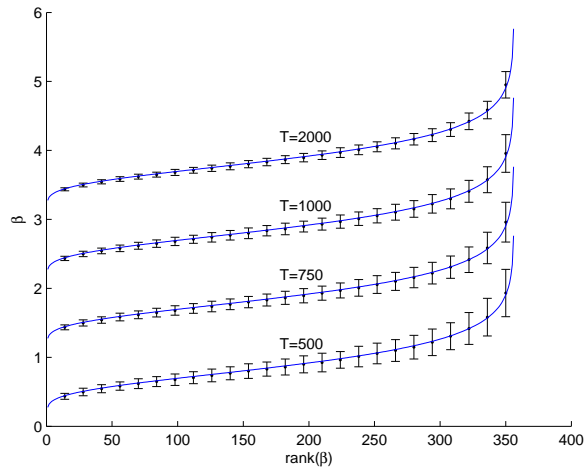}
  \caption{\label{figmc2} Input $\beta$ as function of the rank together with the
           simulated 5\% confidence level error range
           for various window sizes $T$ and $ N=356$. $\gamma_1$ are log-normal
           distributed  with $\gamma_0=0.865$ and $\theta=0.26$. Plotted with an offset of 1 for each series.}
\end{figure} 

We determine $\gamma_{0}$ by optimizing the agreement of the observed
eigenvalue spectrum $N(\lambda)$ with the simulated spectrum. As an example we
show this comparison using $log(1+N(\lambda))$ for the S\&P data in a window of 3
years around 2004 and log-normally distributed $\gamma_{1}$. Both agree surprisingly
well. A Kolmogorov-Smirnov test leads to a p-value of 0.2. In contrast, Laplace
or normally distributed $\gamma_{1i}$ lead to $p\sim 10^{-4}$ or less. The model
accounts apart from the bulk also for the isolated medium eigenvalues attributed 
in the literature to sub-markets \cite{Mant,plerou}. Only the second largest
eigenvalue  attributed to the trading volume \cite{plerou} does not 
correspond to a statistical
fluctuation. 

The result of 200 Monte Carlo repetitions of the dynamic eq. \myref{mc1}
is shown in figure \ref{figmc2} for various $T$. The ordered input $\beta_i$ 
are connected by a line. The errors correspond to a 5\% confidential range for a single 
measurement. There is little dependence on $N$ or the assumed distribution of
$\gamma_1$. In table \ref{tabmc} we give the average error for various $N$ and
different $\gamma_1$ and $T$. The errors only vary with $T$ by $1/\sqrt{T}$.
These simulations prove that within the assumed SVM the values of $\beta_i$ 
can be reliably estimated also for moderate window sizes $T$.

To summarize, for the empirical analysis of $C$ in the next section we make the 
following assumptions: From the market hypothesis we can establish the leading
eigenvector of $C$ as CAPM $\beta$-coefficients. By the SVM assumption the
time average in eq. \myref{aa1} corresponds to an average over the noise. Making
the market dominance assumption the errors on  $\lambda_0$ and $\beta_i$ are
of the order of $1/N,\sqrt{2/T}$.

\section{Transition of the Markets in 2006 \label{ana} }

\begin{figure}[h]
\includegraphics[width=0.95\linewidth, trim = 0 22 0 0 , clip=true]{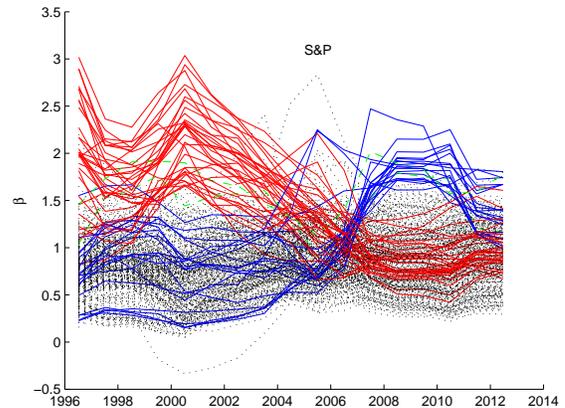}
   \caption{\label{fig1}Time  dependence of $\beta_i$ for 356 stocks of the
            S\&P market. The 35 stocks with largest $\beta$ in 1998-2002 are
            shown in red, the 20 largest in 2007-2010 in blue.}
\end{figure}
\begin{figure}[h]
  \includegraphics[width=0.95\linewidth, trim = 0 22 0 0 , clip=true]{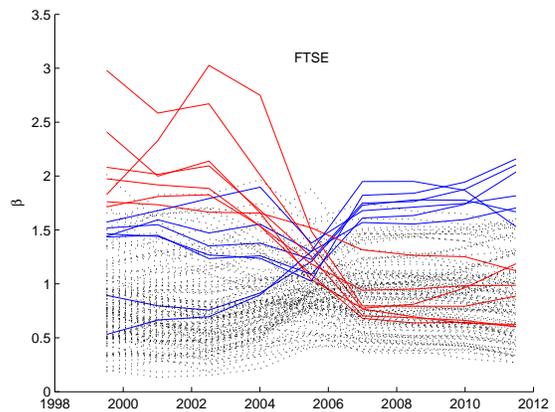}
   \caption{\label{fig2}Time  dependence of $\beta_i$ of the British FTSE market.
            The 7 stocks with largest $\beta$ in 1998-2002 are  shown in red,
            the 7 largest in 2007-2010 in blue.}
\end{figure}
\begin{figure}[h]
\includegraphics[width=0.95\linewidth, trim = 0 22 0 0 , clip=true]{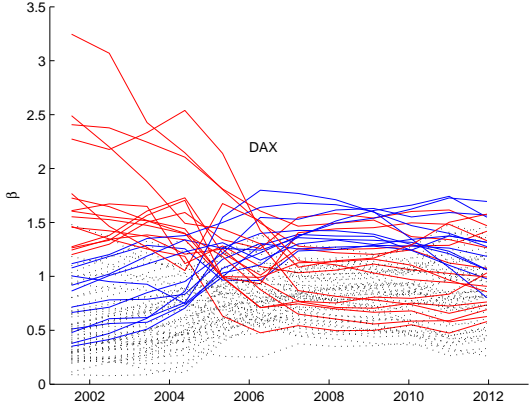}
   \caption{\label{fig3}Time  dependence of $\beta_i$ for 78 stocks of the
            German DAX market.The 15 stocks with largest $\beta$ in 1998-2002 are
            shown in red, the 15 largest in 2007-2010 in blue.}
\end{figure}

We apply our approach to 356 stocks from the
S\&P market, 132 stocks of the British FTSE market, and to 78 stocks from the 
German market. 
 To obtain the possible minimum window size $T$ we look at the large eigenvalue
$\lambda_0(t)$ and the corresponding eigenvector. As the  criterion we use the presence of
(only) positive values of $\beta_i(t)$.
In this way we find  for $T$ a value of roughly 3-4 years for all markets. For a better
visualization of the time variation we use overlapping windows by varying 
$t$ in steps of years.

In figure \ref{fig1} we show the $\beta-$coefficients derived from the largest eigenvector of $C$ for the S\&P market
for a time window of 3 years. Except one case 
around 2001 they are all positive. Some of the stocks exhibit a substantial time variation
with a transition around 2006. Stocks with large $\beta$ during the years
1998-2002 (this time interval is called ITB for IT bubble hereafter) change to small
$\beta$ values around 2006, their values remain low in 2007-2010 (this time interval is called FB
for the finance bubble hereafter). Vice versa those stocks with a large $\beta$ in the
finance bubble exhibit small values before 2006. A similar effect occurs
also for the FTSE market (shown in figure \ref{fig2}) and the German
market (shown in figure \ref{fig3}). For both a window size of 4 years is used. 

\begin{equation} \mylabel{a4}
R(t,s)=A_S \sum_{i \epsilon s}\theta(\beta_i-1.0)\beta_i(t)\; V(t,i)
\end{equation}
The normalization constant $A_S$ is chosen to have $\sum_s\; R(t,s)=1$.

\begin{figure}
\includegraphics[width=\linewidth,trim = 0 20 0 0 , clip=true]{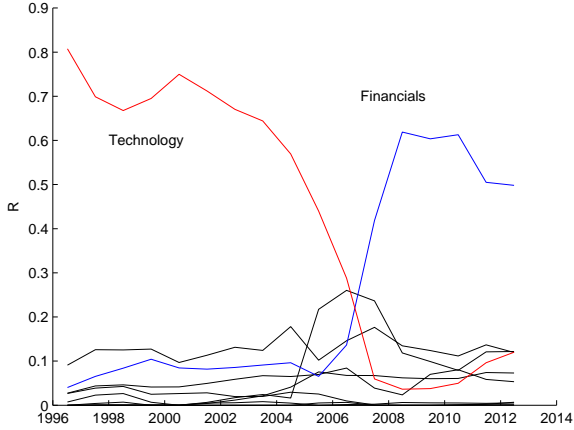}
   \caption{\label{fig4a}Time  dependence of the risk parameter $R(t,s)$ for the
            eight sectors with $R\ne 0$ of the S\&P market.}
\end{figure}
\begin{figure}
\includegraphics[width=\linewidth, trim = 0 20 0 0 , clip=true]{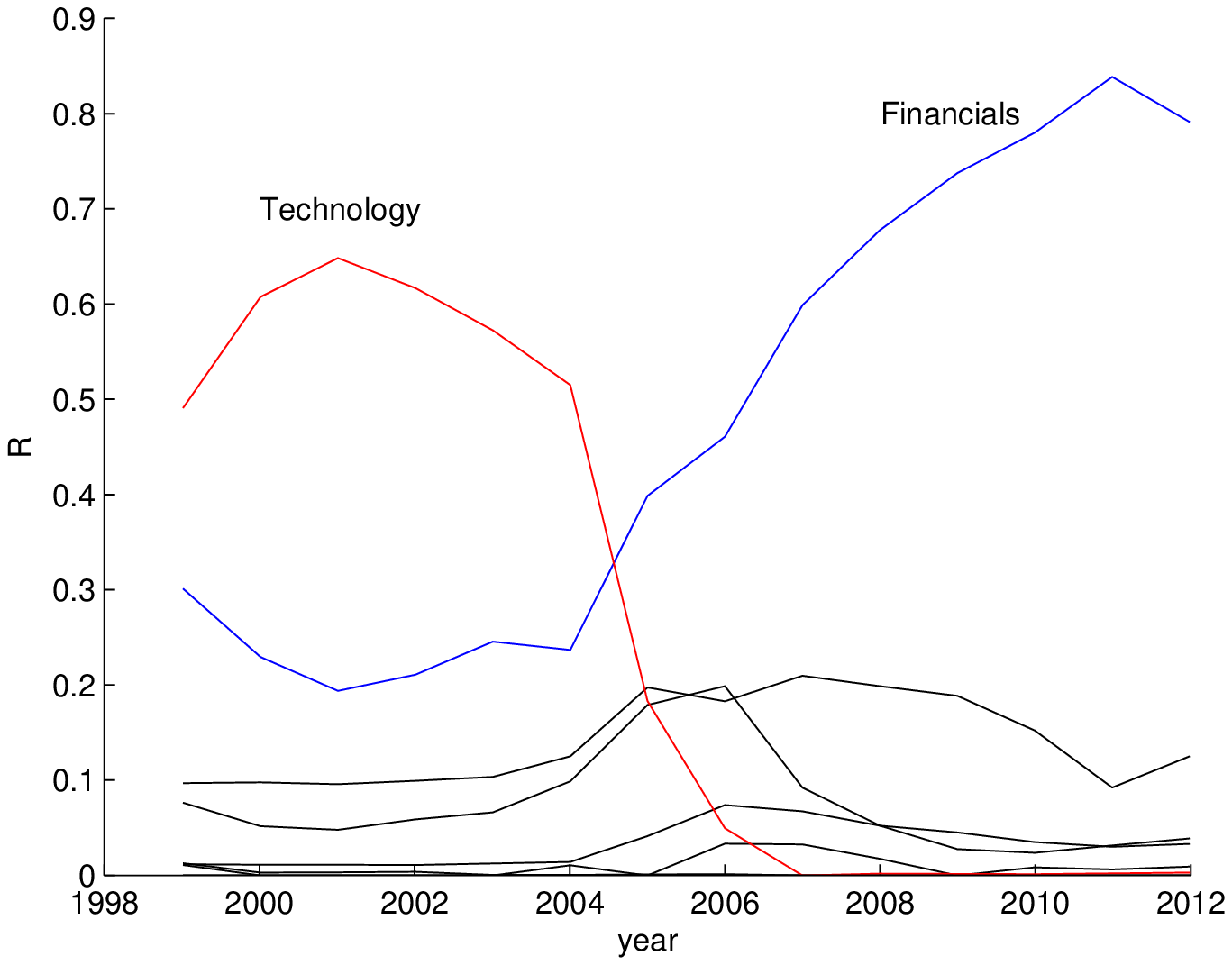}
   \caption{\label{fig4b}Time  dependence of the risk parameter $R(t,s)$ for the
            nine sectors of the FTSE market.}
						\end{figure}
\begin{figure}
\includegraphics[width=\linewidth ,trim = 0 20 0 0 , clip=true]{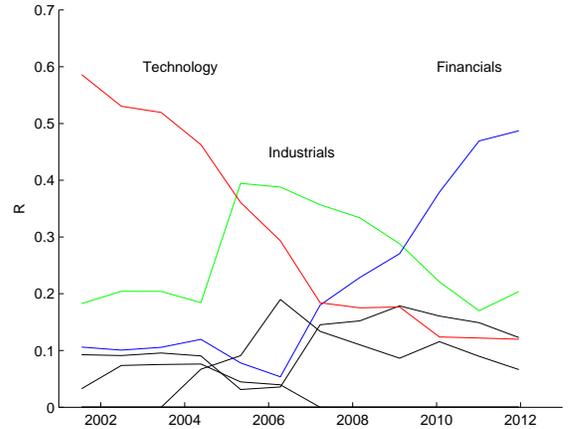}
   \caption{\label{fig4c} Time  dependence of the risk parameter $R(t,s)$ for the
            nine sectors of the DAX market.}
\end{figure}

A more detailed characterization of the market can be obtained
by considering the sector $s$ out of the GICS/TRBS classification for all firms.
An inspection of the firms with large 
$\beta$ during the ITB in figure \ref{fig1} shows that they  dominantly belong
to the IT/technology sector. Likewise firms with large $\beta$ during 
the FB are mostly from the financial sector.
 Since a $\beta>1$ signals a risky investment,
we can define a market risk measure $R(t,s)$ for the sectors by multiplying
$\beta_i>1$ with the number $V(t,i)$ of traded shares in each window. Note that
for the following analysis we merge the sectors IT and telecommunication for the UK and Germany
since we have only few stocks in these sectors and they show similar behavior.

\begin{figure*}[htb!]
\includegraphics[width=0.95\linewidth,bb= 43 323 595 517, clip=true]{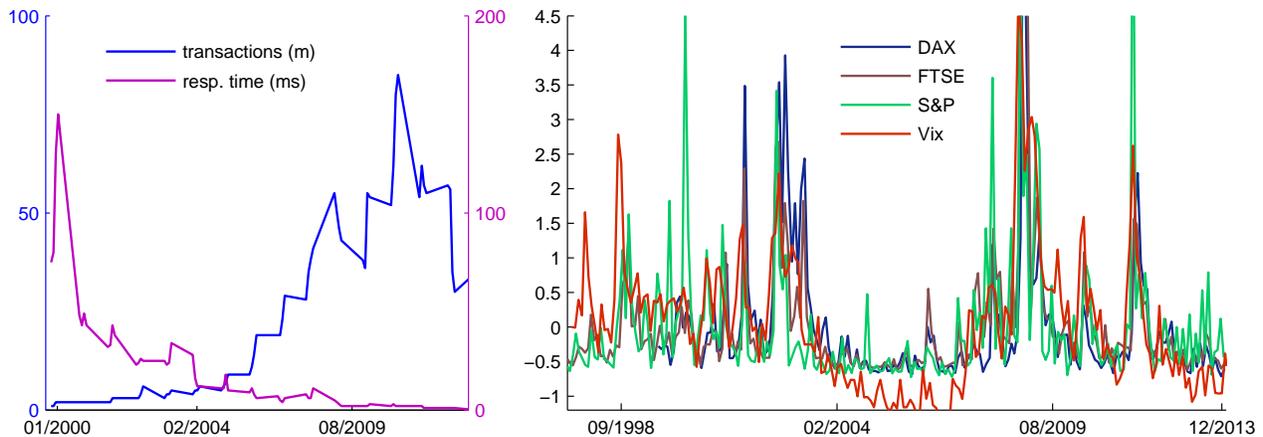}
   \caption{\label{fig5b}(a) Time dependence of response time and traded volume at the Eurex exchange, source: Eurex Exchange \cite{eurex}. (b) Normalized monthly volatility of the S\&P, FTSE and DAX indices, normalized VIX index.}
\end{figure*}

In figure \ref{fig4a} the risk parameters from eq. \myref{a4} for the S\&P market
 is shown as a function
of time. Only the technology sector (red) before the transition in 2006 and the
financial sector (blue) after 2006 exhibit large values of the risk measure. The value of the
risk measure is small for all other sectors. 
Due to the time window of 3 years the time of  the transition
can be fixed only with an error of 1.5 years. A similar phenomenon is seen for the
FTSE market in figure \ref{fig4b} and the German market in figure \ref{fig4c}. 
In contrast to the other markets the industrial sector of the DAX shows a peak
in $R$ at 2006. This may be due to the fact that this sector contains one third of 
all firms. Some are large firms that are difficult to pinpoint to a specific sector. 
Since $R$ is normalized to $\sum_s R(t,s)=1$,
the small $R$ of technology and financials are compensated mainly by the industrial sector.
The transition for the S\&P appears to be
somewhat sharper than for  FTSE and DAX due to the smaller number of stocks in the latter.

\section{Explanations for the observed transition}\label{sec:explain}

\subsection{Interpretation with a diluted Ising model \label{hft}}

\begin{figure}
\includegraphics[width=0.95\linewidth, trim = 20 10 0 0 , clip=true]{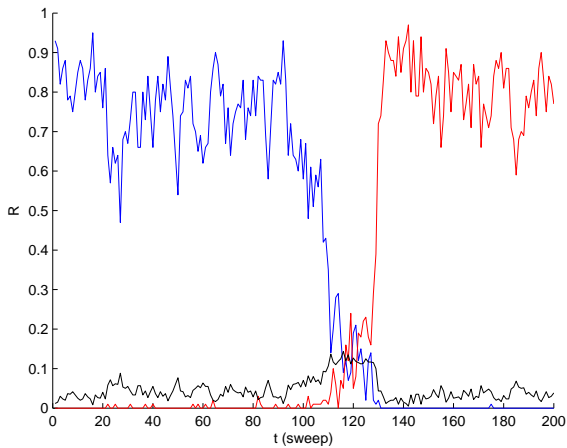}
   \caption{\label{R_simul}Simulation of $R$ as function of time with
             the diluted Ising model. Parameters are $g=3.99$, $A=100$ and 
             $S=8$. For $100\le t\le 120$ a field $h=-0.03$ is applied.
             The black line gives the average of $R(0)$.}
\end{figure}

The behaviour of the risk parameter
$R$ in figures \ref{fig4a}-\ref{fig4c} indicates a phase transition
analogous to models in statistical physics with $R$ as order parameter.
Such a model can be constructed by a generalization of the Ising model.
There are $A$ agents trading one stock per time out of the $S$ sectors.
Each agent $a$ is characterized by a spin value $\sigma_a$. Values $\sigma_a=\pm 1$
 denote trading in the risky sectors IT or financials, and $\sigma_a=0$ denotes
the remaining $S-2$ sectors. We assume that the $\beta$-dependent factor in 
eq. \myref{a4} can be replaced by its mean. Then the normalized $R(s)$ is 
equal  to the fraction of agents  trading in sector $s$. The agents can change 
their opinion due to an interaction between all other agents. At each time 
they chose a new value of $\sigma_a$ by the following probabilities $w(\sigma)$
\begin{align} \mylabel{r1}
w(\pm 1)&=\frac{1}{w_n}\exp(\pm g(m+h)) \\
w(0)&=\frac{1}{w_n} (S-2)
\end{align}
where $m \;A=\sum_a \sigma_a$  and $g$ is the strength of the interaction. $w_n$
normalizes the probabilities. $h$ denotes a possible external field. 
Since agents with $\sigma_a=0$ do not contribute 
to $m$, the model corresponds to a dynamical dilution.
 For small $g$ the system is in the disordered 
state with $R(s)=1/S$ and for large $g$ in the ordered state with
one of the $R(\pm 1)$ becoming large. 

The model can be solved analytically
for large agent numbers $A$,
as shown in appendix \ref{model}. For $S>6$ a first order transition occurs
at a critical value $g_c$. An internal reason for the transition
can be modeled by a dynamically changing $g$. However, this faces the
following problem: To observe a constant $R$ over 6-10 years, like for the S\&P
in figure \ref{fig4a}, the time constant to change $g$ must be in the same 
order of magnitude. In the two years around 2006 the S\&P market changes 
from an ordered state into a disordered and back into another ordered state,
making an external reason more likely. 

Reviewing possible external events around 2006
there seems to be no major political event nor any drastic change in asset 
prices. In fact, the volatility around that time was relatively low, as shown
in the right panel of figure \ref{fig5b}. One observes similarly high
 volatility before and after the transition.
 The only event seems to be the onset of high frequency
trading (HFT) in 2005, see, e.g., \cite{htf2,dark,eurex}. In the left panel of
figure \ref{fig5b} we show (as a representative example) the response time
and the traded volume at the Eurex exchange. The small response time and a maximum 
of the trading volume hint at a growing dominance of
computerized HFT trading after 2005. 

There are several possibilities how HFT
can trigger the transition. Faster trading may reduce the time constant in the
dynamics of $g$.  Generally, computer programs might be more able than
humans in hedging risks. Before 2008 they used almost riskless strategies,
as arbitrage or flash trading, which became less important afterwards.
In the model these effects can be accounted for by a change of $g$ or
the effect of a field $h$. If one chooses $g$ near $g_c$, both phases are
coexistent and only very small changes of $g$ or $h$ are needed to change 
the phase. In figure \ref{R_simul} we show a simulation of $R$ with 
the probabilities from eq. \myref{r1} with constant $g$ and application of a 
small field $h$ at $100 \le t \le 120$, which disfavors a previous risky
sector. Obviously the model can reproduce the observed $R$ for the markets. 
\subsection{High frequency trading and the returns distribution \label{testhft}}

The appearance of HFT should leave traces in the distribution of returns.
Advocates of HFT \cite{eurex} claim that it leads to a more efficient market.
More efficiency should lead to less price changes and therefore to an excess
of smaller returns. Critics \cite{spitzer} assert that computerized trading 
increases instabilities, which amounts to larger returns. Both effects can be 
seen in the pdf for the market return $r_M$. Its pdf can be characterized by the
shape parameter $\alpha$ using the Pareto-Feller parametrization from 
eq. \myref{a3d}. We obtain $\alpha$ by maximizing the Log-Likelihood
$L$ in each window. Errors on $\alpha$ correspond to a change of $L$ by 0.5.

\begin{figure}
\includegraphics[width=0.9\linewidth, trim = 10 22 10 0, clip=true]{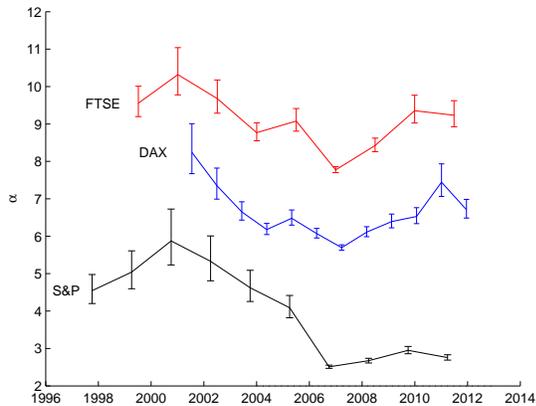}
   \caption{\label{fig5a}Time  dependence of the shape parameter $\alpha$.
            For better readability we added 3 (5) to the values of DAX (FTSE).}
\end{figure}

In figure \ref{fig5a} we show the time dependence of $\alpha$ for the three
markets. Before 2006 one finds values $\alpha \sim 4-5$ with good $\chi^2$
probabilities. For all three markets a drop to values below 3 appears after 2006.
Small $\alpha$ imply a much more enhanced tail of the pdf as expected from HFT.
The $\chi^2$ probabilities are worse in 2006, but
still acceptable on the 5\% level. However, the lower probabilities are due to
systematic deviations from \myref{a3d}. 

\begin{figure}[ht]
\includegraphics[width=\linewidth]{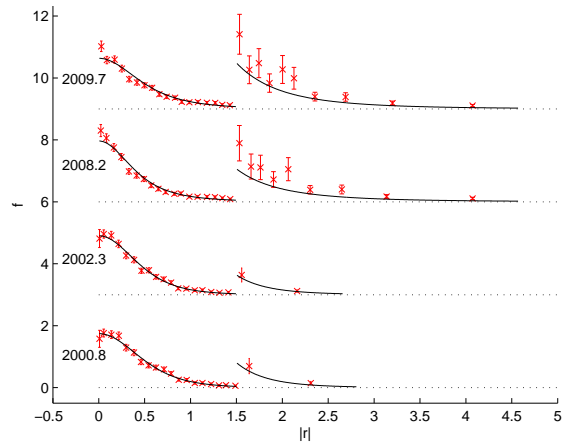}
   \caption{\label{fig6c} Distribution $f$ of the standardized ($<r_M^2>=1$) market
           return $|r_M|$ for the S\&P market. For $|r_M>1.5|$ $f$ is multiplied with 20.}
\end{figure}

In figure \ref{fig6c} we show some typical
pdfs of the market returns before and after 2006 for the S\&P market.
We see a perfect description by eq. \myref{a3d} for the returns from the time
windows centered in 2000 and 2002, whereas for those centered in
2008 and 2009 a substantial excess at $r\sim 0$ occurs, and the badly described
tail extend to much larger value as before. This behavior is expected from HFT.

For the European markets, shown in figures \ref{fig6b} and \ref{fig6a} in the appendix,
the excess of small returns is less significant.
Only a significantly enlarged tail is observed after 2006. These markets
might be less affected by HFT and therefore only the instability effect is seen.
There is no trace in the volatility, since both effects can cancel out in $E[r^2]$.

\section{Conclusions}\label{sec:conc}

The literature on regularities in asset returns has for a long time argued that 
the $\beta$ values of stocks are time varying. We have shown that we can extend the
concept of $\beta$ values to systematically describe a risk measure for stocks from 
different sectors of the economy. This slowly varying sector
specific risk measure describes ordered states in the market and identifies sectors 
which show concentration of market risk. A possible trigger for the observed transition may
be the onset of high frequency trading in 2005.

\begin{appendix}

\section{Perturbation Theory \label{pert}}
Assume a matrix $C$ can be written as $C=C_0+C_1$ with a small perturbation $C_1$.
$C_0$ has one large eigenvalue $E_0$ and $N-1$ degenerate zero eigenvalues. Due 
to the degeneracy we can impose for the eigenvectors $e_i^\mu$ with $\mu>0$
of $C_0$ the conditions
\begin{equation} \mylabel{A1}
e^\nu \cdot (C_1\; e^\mu)=0 \quad \mbox{for} \quad \mu,\nu >0,\nu \ne \mu
\end{equation}
The eigenvalues $\lambda_\mu$ and eigenvectors $f_i^\mu$ of $C$ can be expanded
in a power serie in $C_1/E_0$. For $\mu=0$ we get
\begin{equation} \mylabel{A2}
\lambda_0=E_0+e^0 \cdot (C_1\;e^0)+\frac{1}{E_0} \left[ e^0 \cdot (C_1^2\;e^0)- \left( e^0 \cdot (C_1\;e^0) \right)^2 \right]
\end{equation}
\begin{equation} \mylabel{A3}
f_i^0=\left [1-\frac{1}{E_0} e^0 \cdot (C_1\;e^0)\right ]e^0_i +  \frac{1}{E_0}(C_1\;e^0)_i
\end{equation}
The remaining eigenvectors need the solution of condition \myref{A1}
\begin{equation} \mylabel{A4}
\lambda_\mu= e^\mu \cdot (C_1\;e^\mu)-\frac{1}{E_0} \left(e^0 \cdot (C_1\;e^\mu) \right)^2
\end{equation}
\begin{equation} \mylabel{A5}
f_i^\mu=e_i^\mu-\frac{1}{E_0}(e^0 \cdot (C_1\;e^\mu) e_i^\mu
\end{equation}
Note that this expansion reproduces the exact result for $tr\; C$ and for $C_1$ proportional 
to a unit matrix.\\
With $(C_0)_{ij}=\gamma_i\gamma_j$ we have $E_0=(\gamma,\gamma)=\gamma^2$ and
$e^0_i=\gamma_i/\sqrt{\gamma^2} $. Inserting
$(\gamma,C_1^k\gamma)=A_k\;\gamma^2$ into eqns. \myref{A2} and \myref{A3} we get for
$\lambda_0$ and $\beta_i=\sqrt{N}f_i^0$
\begin{equation} \mylabel{A6}
\lambda_0=\gamma^2+A_1+\frac{1}{\gamma^2}(A_2-A_1^2)
\end{equation}
\begin{equation} \mylabel{A7}
\beta_i=\left [1-\frac{1}{\gamma^2}A_1\right ]\sqrt{\frac{N}{\gamma^2}}\gamma_i
            +\frac{1}{\gamma^2}a_i
\end{equation}
with $a^2=(N/\gamma^2)A_2$. Market dominance implies $E_0=\gamma^2\propto N$ and the 
constants $A_k$ are of order 1. Eqns. \myref{A6} and \myref{A7} show that up to 
corrections of order $1/N$ the leading eigenvalue $\lambda_0$ and its eigenvector 
$\beta_i$ do not depend on $C_1$.

\section{Diluted Ising Model \label{model}}
The transition probabilities \myref{r1} correspond to the heat bath algorithm
for the following equilibrium distribution
\begin{equation} \mylabel{B1}
w(\sigma)=\frac{1}{Z}\exp\left[ A\cdot g( m(\sigma)^2/2+hm(\sigma))\right]
\end{equation}
with $m(\sigma)=(1/A)\sum_a \sigma_a$.
The partitioned sum $Z$ we calculate by 
using the Gaus trick $\sqrt{\pi}\exp(m^2)=\int dx \exp(-x^2+2mx)$ and
evaluating the integral for large $A$. We get for $Z$
\begin{equation} \mylabel{B2}
\ln Z=A\left[ \ln (S-2+2ch((m_0+h)g)) - \frac{g}{2} m_0^2 \right]
\end{equation}
The expectation value $m_0$ of $m(\sigma)$ must maximize $\ln Z$. This
leads to the so called mean field condition for the order parameter $m_0$
\begin{equation} \mylabel{B3}
m_0=\frac{2sh(m_0(g+h))}{S-2+2ch(m_0(g+h))}
\end{equation}
The fraction $R_{\pm}$ at $h=0$ of agents with $\sigma_a=\pm 1$ is given by
\begin{equation} \mylabel{B4}
R_{\pm}=\frac{\exp(\pm m_0g)}{S-2+2ch(m_0g)}
\end{equation}
Eqns. \myref{B2} and \myref{B3} lead already at $h=0$ to a surprisingly 
rich spectrum of
phases depending on the number of sectors $S$ and $g$.
For $S<6$ we have a similar behaviour as in the Ising model ($S=2$).
At $g_c=S/2$ a second order transition occurs. At $S=6$ the 
transition is still of second order, but with different critical exponents.
\begin{figure}[h]
\includegraphics[width=0.7\linewidth]{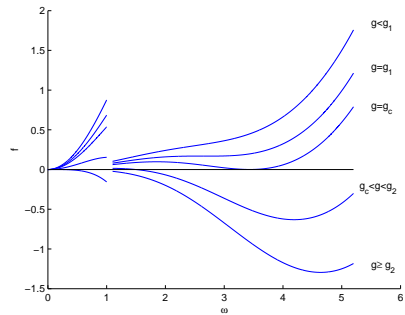}
   \caption{\label{free}$-\ln Z/(gA)$ as function of $\omega=gm_0$ for
            various values of $g$. Values for $\omega<1$ are multiplied
            with 10.}
\end{figure}
The general case $S>6$ is illustrated by a plot of $-\ln Z/(gA)$ in figure
\ref{free}. Below $g<g_1$ only the disordered phase exists. At $g=g_1$ a
$m_0\ne 0$ solution appears corresponding a metastable ordered phase, 
since $\ln Z\le \ln Z(m_0=0)$. At $g=g_c$ both phases coexist. 
For $g>g_c$ the $m_0=0$ phase becomes metastable.
Finally for $g>g_{2}=S/2$ only the $m_0\ne 0$ solution exists.
The critical values of $g$ and the values of $R(1)$ at criticality 
corresponding to a lower bound are given in table \ref{tabmodel}.
\begin{table}
\begin{center}
\begin{tabular}{|c |c| c| c | c|}
\hline
 $S$ & $g_{1}$ & $g_c$ & $g_{2}$ & $ R(g_c)$ \\ \hline
  6  &    -    & 3     & -          & 1/6  \\
  8  &  3.73   & 3.82  & 4          & 0.70  \\
  9  &  3.97   & 4.20  & 9/2        & 0.81  \\
  10 &  4.19   & 4.58  & 5          & 0.87 \\ 
 \hline
\end{tabular}
\end{center}
\caption{\label{tabmodel} Critical values of $g$ and $R(g_c)$ at $g=g_c$.} 
\end{table}

\newpage
\section{Return distributions for the European markets \label{morepdfs}}

\begin{figure}[h]
\includegraphics[width=\linewidth]{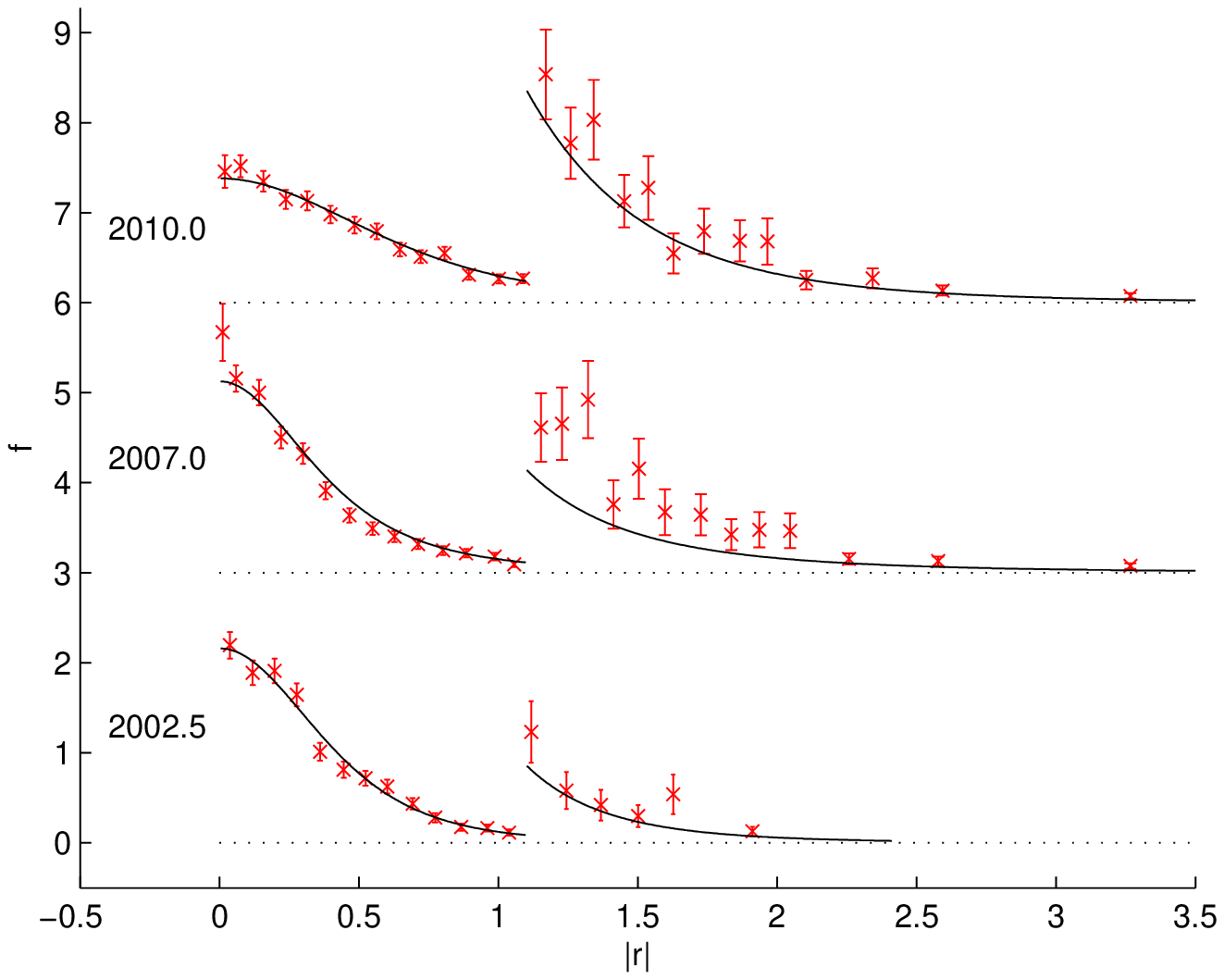}
  \caption{\label{fig6b} Distribution of $f$ for the standardized ($<r_M^2>=1$) market
           return $|r_M|$ for the FTSE market. For $|r_M>1.1|$ $f$ is multiplied with 10.}
\end{figure}
\begin{figure}[h]
\includegraphics[width=\linewidth]{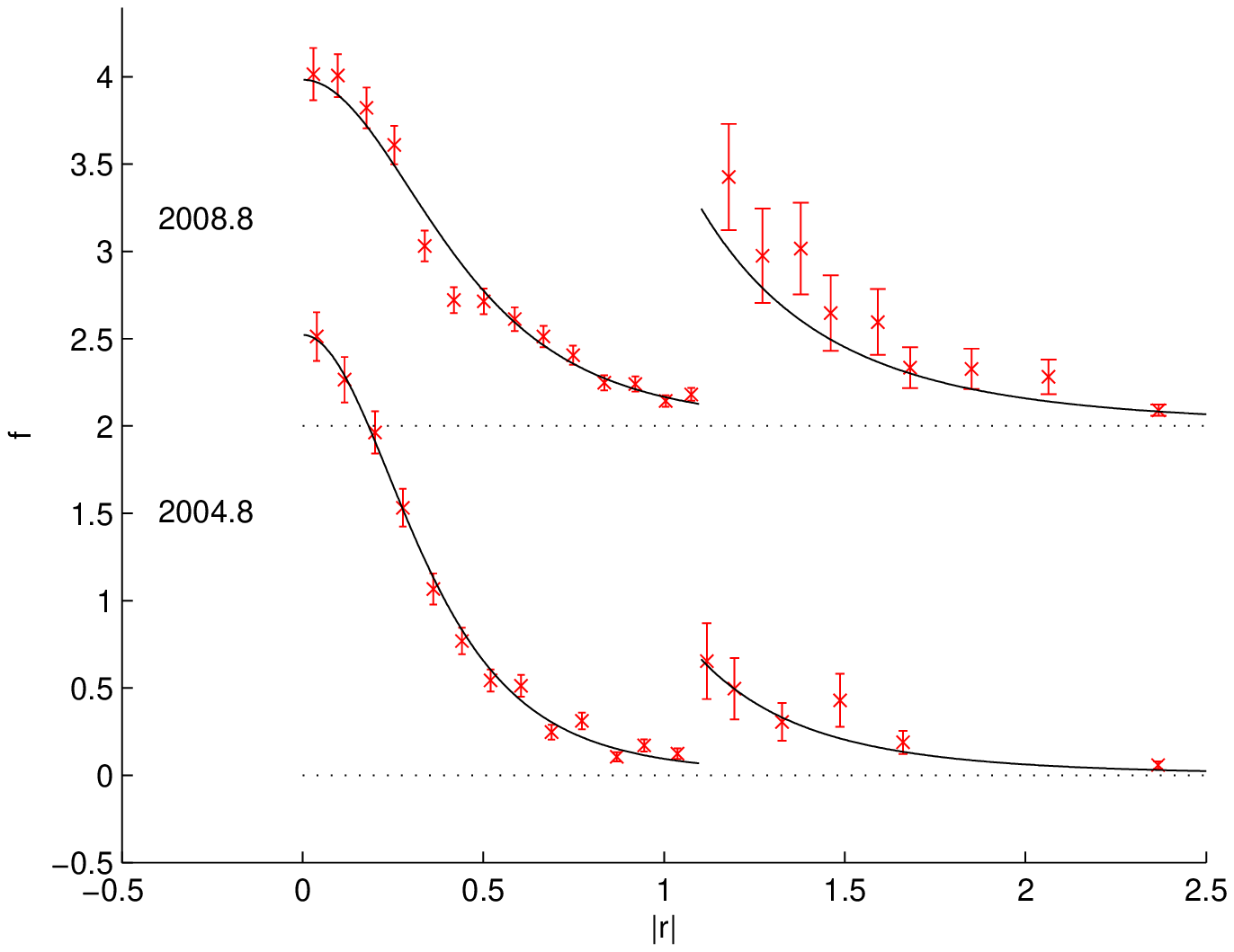}
 \caption{\label{fig6a} Distribution of $f$ for the standardized ($<r_M^2>=1$) market
           return $|r_M|$ for the DAX market. For $|r_M>1.1|$ $f$ is multiplied with 10.}
\end{figure}

\clearpage
\end{appendix}

\end{document}